\documentclass[pra,aps,amsmath,amssymb,onecolumn]{revtex4}

\usepackage{epsfig}

\usepackage{bm}

\usepackage{color}

\begin{document}

\preprint{}


\newcommand{\stef}[2]{$\blacktriangleright${\sc round #1:}{\em #2}$\blacktriangleleft$}

\title{Multiple quantum NMR dynamics in a gas of spin-carrying molecules in fluctuating nanopores}

\author{E.B.Fel'dman}

\email{efeldman@icp.ac.ru}

\author{A.I.Zenchuk}

\email{zenchuk@itp.ac.ru}

\affiliation{Theoretical Department of Institute of Problems of  Chemical Physics of the Russian Academy of Sciences, Chernogolovka, Moscow Region, 142432, Russia}


\begin{abstract}

The effect of Gaussian fluctuations of nanopores filled with a gas of spin-carrying molecules ($s=1/2$) on the multiple quantum (MQ) NMR dynamics is investigated at different variances and correlation times of the fluctuations. We show that the fluctuations smooth out the evolution of  MQ NMR coherence intensities which rapidly oscillate as functions of time in the absence of fluctuations.
The  growth and decay  of the MQ coherence clusters in the fluctuating nanopore are also investigated. 

\end{abstract}


\maketitle

\section{Introduction}

{ The} multiple quantum (MQ) NMR \cite{BMGP} is not only a powerful tool for the investigation of  nuclear spin distributions in solids but also an effective method to study  MQ NMR coherence relaxation in large systems of highly correlated spins \cite{AS,AS2}. With this tool, one can estimate the decoherence time  \cite{KS} and a possible distance of { the} quantum information  transfer \cite{AS}. These parameters of quantum systems are  important for  quantum information processing \cite{NC}.

However, dipolar coupling constants in  real physical spin systems do not remain time-independent because of such effects as 
 molecular motions and imperfect experimental realizations of the two-spin/two-quantum Hamiltonian  \cite{BMGP} together with  the effect of high order corrections \cite{DFFZ} to the average Hamiltonian theory \cite{HW}. All these effects  lead to uncontrolled  variations of the dipolar coupling constants. 
 A fluctuating nanopore filled with a gas of spin-carrying ($s=1/2$) molecules (atoms) is 
a suitable model for the study of the MQ NMR dynamics with { variable} dipolar coupling constants  \cite{FR}.  This study is stimulated by experimental results concerning fluctuations in  nanostructures.
 For example, fluctuating nanostructures emerge in carbon nanotubes \cite{PWUH} driven by mechanical or electrical excitation. Other examples include gas vesicles \cite{SBBGH} and nanobubbles of insoluble gas in a liquid \cite{KNTSO}.  We expect that  the results obtained in our paper can be useful for experimental investigations of fluctuations in nanopore materials. Because of  the nanopore fluctuations, the  dipolar coupling constant,
averaged over molecular motion, fluctuates as well.  Since the characteristic time of molecular movements is much less than the spin flip-flop times, the averaged dipolar coupling constant  is the same for all spin pairs \cite{BKHWW,FR2} even in fluctuating nanopores. { Jeener has recently shown \cite{J} that the mean field approach involving the distant dipolar field \cite{J1} leads to the results which are analogous to those obtained in \cite{BKHWW,FR2}, if equilibrium fluctuations
are taken into account  in the estimation of the dipolar field. The nuclear spin dynamics and the NMR line shape of a  nuclear spin system placed in a fluctuating nanocontainer was investigated in \cite{KN,KSN} on the basis of the method developed in Ref.\cite{FR}.}

In the present paper we study  the MQ NMR dynamics of spin-carrying ($s=1/2$) molecules (atoms) of a gas in the fluctuating nanopores. 
We investigate  the relaxation (decoherence) of the MQ coherence intensities 
using the method developed in  
Refs. \cite{AS,AS2}  and applying the theoretical approach explored in Refs. \cite{FR,DFFZ2,DFZ}. The latter  allows us  to study the MQ NMR spin dynamics in a   system of hundreds of equivalent spins.    In particular, it becomes possible  to investigate the growth of the MQ coherence clusters in a fluctuating nanopore. Assuming Gaussian character of nanopore fluctuations we can investigate the MQ NMR dynamics at different variances  and correlation times of the fluctuations.

The paper is organized as follows. The theoretical approach to MQ NMR dynamics in the fluctuating nanopores is developed in Sec.\ref{Section:theory}.
The numerical simulation of the  evolution of MQ NMR coherences  in the system  of 201 equivalent spins as well as  the evolution of the MQ NMR coherence clusters  is presented in Sec.\ref{Section:numerics}. We briefly summarize our results and discuss further perspectives in the concluding Sec.\ref{Section:conclusions}.


\section{Intensities of MQ NMR coherences in  fluctuating nanopores}

\label{Section:theory}

\begin{figure*}
   \epsfig{file=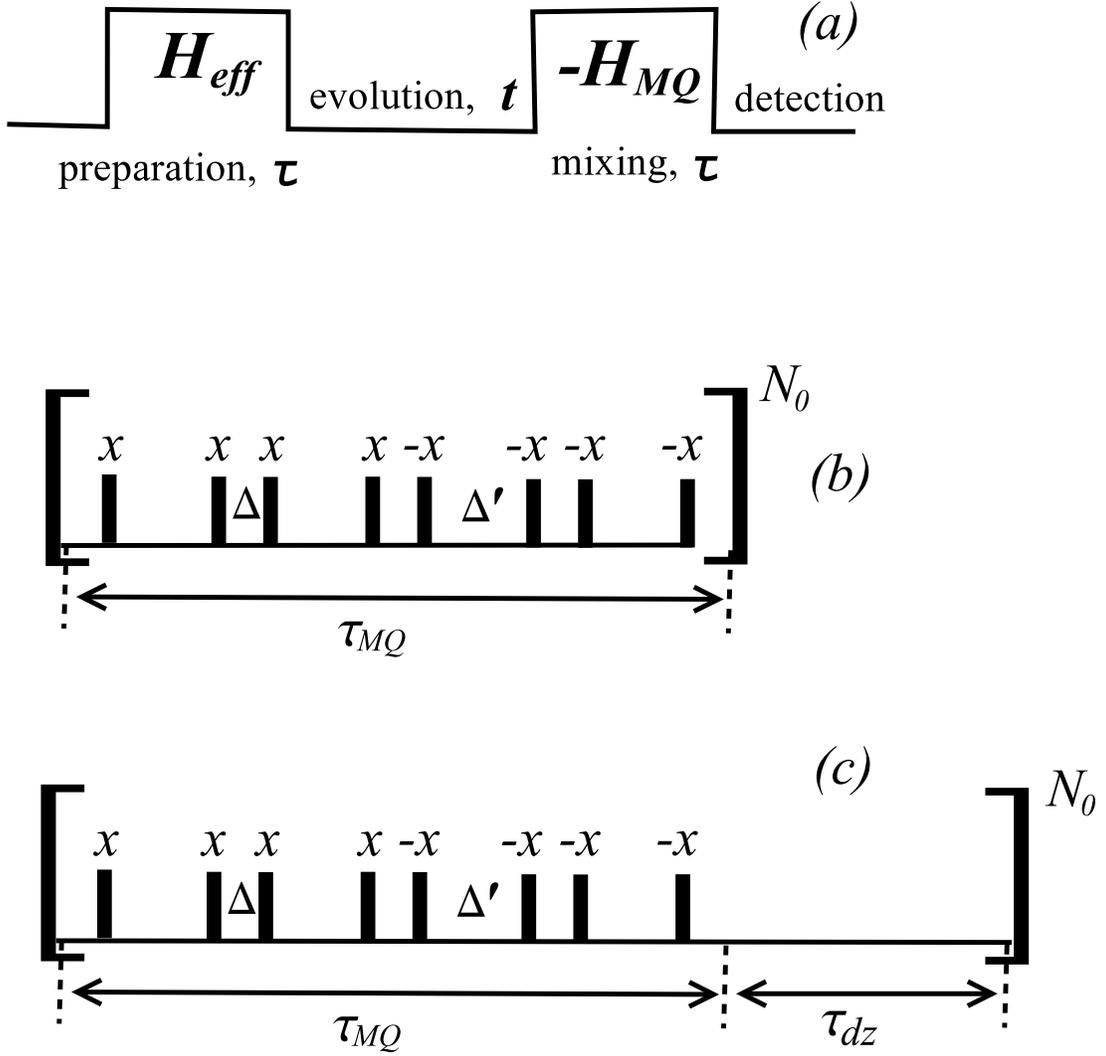,scale=0.65,angle=0}
\caption{$a$) The basic scheme of the MQ NMR experiment. The Hamiltonians $H_{eff}$ (\ref{Heff0}) and $H_{MQ}$ (\ref{HMQ}) govern the spin dynamics on the  preparation and mixing periods of the MQ NMR experiment respectively. $b$) The  $\pi/2$
pulse sequence \cite{BMGP}. $\Delta$ and $\Delta'=2 \Delta+\tau_p$ are time intervals between pulses, $\tau_p$ is the $\pi/2$ pulse duration. The period of the pulse sequence is $\tau_{MQ}$.
The preparation period  is formed by  $N_0$ cycles of the above pulse sequence,  thus $\tau=N_0\tau_{MQ}$, $H_{eff}=H_{MQ}$, see Eq.(\ref{HMQ}). $c$) The modified pulse sequence \cite{AS} on the preparation period. The period of pulse sequence is $\tau_{MQ}+\tau_{dz}$, thus $\tau=N_0 (\tau_{MQ}+\tau_{dz})$. The  Hamiltonian $H_{eff}$ is given by Eq.(\ref{Heff0}) }
 \label{Fig:0}
\end{figure*}

We consider the MQ NMR experiment with a gas  of $N$ spin-carrying ($s=1/2$) particles  in a fluctuating nanopore which is  placed  in the strong external magnetic field
$\vec H_0$ directed along $z$-axis  \cite{BMGP}. The MQ NMR experiment consists of four distinct periods of time (see Fig.\ref{Fig:0}$a$): preparation ($\tau$), evolution ($t$), mixing ($\tau$) and detection. On the preparation period, the spin system is irradiated by the proper multipulse sequence of radio-frequency pulses, rotating spins by  $\pi/2$ around  the $x$ axis (so-called $\pi/2$ pulses), see Fig.\ref{Fig:0}$b,c$. As a result, the dynamics of  the spin system is described by the effective Hamiltonian $H_{eff}$ in the rotating reference frame (Fig.\ref{Fig:0}$a$).  We consider two types of pulse sequences on the preparation period depicted in Fig.\ref{Fig:0}$b$ and $c$.  The first one (Fig.\ref{Fig:0}$b$) is the standard pulse sequence resulting in the averaged nonsecular two-spin/two-quantum Hamiltonina $H_{MQ}$ \cite{BMGP}, i.e. $H_{eff}=H_{MQ}$.
The second pulse sequence was introduced in \cite{AS}, see Fig.\ref{Fig:0}$c$. It yields the modified expression for $H_{eff}$ which  is explored below.

 Our subsequent results are based on two assumptions: ($i$)  the characteristic time of the nanopore fluctuations is much longer than that of molecular motion in the nanopore and ($ii$) the characteristic time of fluctuations is also much longer than the period of the pulse sequence in Fig.\ref{Fig:0}$b$.  Then the MQ NMR dynamics on the preparation period is governed by the following  two-spin/two-quantum Hamiltonian $H_{MQ}$ \cite{DFFZ2}:
\begin{eqnarray}\label{HMQ}
H_{MQ}=-\frac{D(t)}{4} \Big((I^+)^2 + (I^-)^2\Big),
\end{eqnarray}
where
 $I^+=I_x+iI_y$ and $I^-=I_x-iI_y$ are the total raising and lowering operators, $I_\alpha$ ($\alpha=x,y,z$) is the projection operator of the spin angular momentum on the axis $\alpha$ and $D(t)$ is the dipolar coupling constant in the fluctuating nanopore,
\begin{eqnarray}\label{Dt0}
D(t)=\gamma^2 \hbar \frac{f(t)}{V(t)} ( 3 \cos^2 \theta(t) -1).
\end{eqnarray}
Here $\gamma$ is the gyromagnetic ratio, $V$ is the volume of the nanopore, $f$ is the form-factor, depending on the shape of the nanopore, and $\theta$ is the angle determining the orientation of the nanopore with respect to the external magnetic field  \cite{BKHWW,FR2}. The  explicit expression for the form-factor $f$ associated with the  ellipsoidal nanopore  can be found in Ref. \cite{FR2}. Eq. (\ref{Dt0}) shows that the value of the coupling constant significantly depends on the orientation of the nanopore and vanishes at the so-called ''magic'' angle   $\theta_m$ \cite{EBW} defined as
$ \cos \theta_m =1/\sqrt{3}$. 
We also need the Hamiltonian, responsible for secular dipole-dipole interactions in the fluctuating nanopore \cite{FR2},
\begin{eqnarray}\label{Hdz}
 H_{dz}=\frac{D(t)}{2}(3 I_z^2 - I^2),
\end{eqnarray}
and the expression for the Hamiltonian
$H_{eff}$, governing the spin dynamics on the modified  preparation period (cf. Ref.\cite{AS}; see Fig.\ref{Fig:0}$c$),
\begin{eqnarray}\label{Heff0}
H_{eff}=(1-p) H_{MQ} + p H_{dz},
\end{eqnarray}
where  $p=\tau_{dz}/(\tau_{dz}+\tau_{MQ})\le 1$.

For a fluctuating nanopore, we may write
\begin{eqnarray}\label{Dt}
D(t) = \langle D \rangle + \delta D(t).
\end{eqnarray}
Here $\langle D \rangle$ is the average dipolar coupling constant while  $\delta D(t)$ is its fluctuation.
Hereafter we assume the Gaussian fluctuations of the coupling constant $D(t)$ \cite{FR} and characterize
the Gaussian random noise, $\delta D(t)$, by the first two moments
\begin{eqnarray}\label{mom}
\langle \delta D(t)\rangle =0, \;\;\langle \delta D(t_1) \delta D(t_2) \rangle =\langle (\delta D)^2 \rangle C(|t_1-t_2|). 
\end{eqnarray}
Here $\langle (\delta D)^2 \rangle$ is the  fluctuation variance and $C(t)$ denotes the correlation function. In many cases \cite{A}
\begin{eqnarray}\label{C}
C(t)=\exp(-t/\tau_c)
\end{eqnarray} 
where $\tau_c$ is the correlation time.

An important property of the Hamiltonian $H_{MQ}$ (\ref{HMQ}) is that it 
commutes with the operator of the square of the total spin angular momentum $ I^2$ \cite{DFFZ2}. Since the operator $ I^2$ commutes with  $I_z$, it is suitable to study the MQ NMR dynamics in a nanopore using the basis of common eigenvectors  of the operators $ I^2$ and $I_z$ \cite{DFFZ2,DFFZ}.  This choice of the basis allows us to split the problem into a set of simpler problems  for different values of $ I^2$. This resolves the problem of  the Hilbert space dimension growth 
with the increase  in the number of spins. Owing to this simplification it becomes possible to investigate the MQ NMR dynamics in systems consisting of several hundreds of spins and to calculate the profiles of the intensities of MQ NMR coherences (the intensities of MQ NMR coherences versus their order) \cite{DFFZ2,DFFZ}. {  Eq.(\ref{HMQ}) shows that the same simplification is also valid for a fluctuating nanopore.}

In order to investigate the MQ NMR dynamics, one should first find the density matrix $\rho(t)$ solving the Liouville evolution equation \cite{G}
\begin{eqnarray}\label{Liouville}\label{L1}
i\frac{d\rho}{dt} =\left[
-\frac{1}{4} D(t) \Big((I^+)^2+(I^-)^2\Big),\rho(t)
\right]
\end{eqnarray}
with the initial density matrix  $\displaystyle\rho(0)\approx \frac{1}{2^N}(1+\frac{\hbar \omega_0}{kT}I_z)$ written for  the high temperature approximation \cite{G}. Here  $k$ is the Boltsman constant, $T$ is the temperature and $\omega_0=\gamma H_0$. Introducing the new variable 

\begin{eqnarray}
\label{phi}
\varphi(t)=\frac{1}{\langle D \rangle }\int_0^t D(t') dt',
\end{eqnarray}
one can rewrite Eq.(\ref{Liouville}) as follows:
\begin{eqnarray}
\label{L2}
&&
i\frac{d\rho(\varphi)}{d\varphi}=[\bar H_{MQ},\rho(\varphi)],\\\label{bHMQ}
&&
\bar H_{MQ}= 
-\frac{\langle D \rangle }{4} \{(I^+)^2+(I^-)^2\},
\end{eqnarray}
which is 
the usual equation (up to some notations)  for the MQ NMR dynamics. The solution of Eq.(\ref{L2})  was studied numerically in \cite{DFFZ2,DFFZ}. 
A similar consideration may be given to the evolution of the density matrix under the secular  Hamiltonian $H_{dz}$.
Following the results of Refs.\cite{DFFZ2,DFFZ} we represent the solution of Eq.(\ref{L2}),
\begin{eqnarray}
\rho(\varphi)= e^{-i\varphi \bar H_{MQ}} I_z 
e^{i\varphi \bar H_{MQ}},
\end{eqnarray}
as the sum of the contributions $\rho_k$ responsible for the MQ NMR coherences of different orders $k$  ($|k|\le N$)
\begin{eqnarray}
\rho(\varphi)
=\sum_k\rho_k(\varphi).
\end{eqnarray}
The intensities $J_k(\varphi)$ ($|k|\le N$) of the MQ NMR coherences were found  in Ref.\cite{DFFZ2}:
\begin{eqnarray}\label{J1}
J_k(\varphi)=\frac{{\mbox{Tr}}\{
\rho_k(\varphi)\rho_{-k}(\varphi)
\}}{{\mbox{Tr}}(I_z^2)}.
\end{eqnarray}
 We also consider the decay of the MQ  coherence intensities in the MQ NMR experiment with  the modified preparation  period  \cite{AS} (see Fig.\ref{Fig:0}$c$) concatenating the short evolution periods $\tau_{dz}$  under the averaged over fluctuations  secular dipole-dipole Hamiltonian (\ref{Hdz})
\begin{eqnarray}
\bar H_{dz}=\frac{\langle D\rangle}{2}(3 I_z^2 - I^2)
\end{eqnarray}
(which is considered as a  perturbation) with the evolution period $\tau_{MQ}$ under the nonsecular average two-spin/two-quantum Hamiltonian
$\bar H_{MQ}$ (\ref{bHMQ}).
Then the the Hamiltonian $H_{eff}$ (\ref{Heff0})  averaged over fluctuations is \cite{AS}
\begin{eqnarray}\label{Heff}
&&
\bar H_{eff}(p)=(1-p) \bar H_{MQ} + p \bar H_{dz}.
\end{eqnarray}
In this case, the intensities of the MQ coherences $J_k(\varphi,p)$ are the following \cite{DFZ}:
\begin{eqnarray}\label{J2}
J_k(\varphi,p)=\frac{{\mbox{Tr}}\{
\tilde \rho_k(\varphi,p)\rho_{-k}(\varphi,p)
\}}{{\mbox{Tr}}(I_z^2)}.
\end{eqnarray}
Here we use the following representation for the density matrix $\tilde \rho$:
\begin{eqnarray}
\;\;\;\tilde \rho(\varphi,p) = e^{-i\varphi \bar H_{eff}} I_z 
e^{i\varphi \bar H_{eff}} 
=\sum_k\tilde \rho_k(\varphi),
\end{eqnarray}
where  $\tilde \rho_k$ is the contribution to $\tilde \rho$ 
from the MQ coherence of the $k$th order.
 We will use both  intensities (\ref{J1}) and (\ref{J2})  in our calculations.

It is suitable to write 
\begin{eqnarray}
\label{J_k}
J_k(\varphi) = J_k(t+X(t)),
\end{eqnarray}
where 
\begin{eqnarray}
X(t)=\frac{1}{\langle D\rangle}\int_0^t \delta D(t') dt'.
\end{eqnarray}
By the definition of  the Dirac delta function one has the following identity:
\begin{eqnarray}\label{J_kX}
J_k(t+X)=\int_{-\infty}^\infty J_k(t')\delta(t'-t-X(t)) dt'.
\end{eqnarray}
Using the integral representation of the $\delta$-function,
\begin{eqnarray}\label{delta}
\delta(x) =\frac{1}{2\pi}\int_{-\infty}^\infty d\omega e^{-i\omega x},
\end{eqnarray}
we transform Eq.(\ref{J_kX}) to the following one:
\begin{eqnarray}\label{J_kX2}
J_k(t+X)=\frac{1}{2\pi}\int_{-\infty}^\infty d\omega  J_k(t') e^{-i\omega (t'-t)} e^{i\omega X(t)}dt'.
\end{eqnarray}
This equation can be easily averaged over fluctuations. First of all, averaging  both sides of Eq.(\ref{J_kX2})  one can write:
\begin{eqnarray}\label{J_kX2avr}
\langle J_k(t+X(t)) \rangle=\frac{1}{2\pi}\int_{-\infty}^\infty d\omega  J_k(t') e^{-i\omega (t'-t)}
\langle e^{i\omega X(t)}\rangle dt'.
\end{eqnarray}
It is known that  \cite{A,FR} 
\begin{eqnarray}\label{Exp}
\langle e^{i\omega X(t)}\rangle =\exp\Big(-\omega^2 \langle (\delta D)^2\rangle T^2(t)\Big)
\end{eqnarray}
for the Gaussian fluctuations, where 
\begin{eqnarray}
\label{T}
T^2(t)=\int_0^t (t-t')C(t') dt'.
\end{eqnarray}
Substituting Eqs.(\ref{Exp}) and (\ref{T}) into Eq. (\ref{J_kX2avr})  one can obtain after simple calculations:
\begin{eqnarray}\label{J_kX2avr_tr}
\langle J_k(t+X(t)) \rangle=\frac{1}{2\sqrt{\pi \langle (\delta D)^2\rangle T^2(t)}}\int_{-\infty}^\infty dt'  J_k(t')
 \exp(-\frac{(t'-t)^2}{4 \langle (\delta D)^2\rangle T^2(t)} ).
\end{eqnarray}
Eq.(\ref{J_kX2avr_tr}) will be used in Sec.\ref{Section:numerics} for study of the dependencies of the  MQ NMR coherence intensities  on different variances and correlation times of the fluctuations in the system consisting of 201 spins. 
Here the intensities $J_k$ are defined by one of the formulas (\ref{J1}) or (\ref{J2}) and 
the correlation function, $C(t)$, is taken from Eq.(\ref{C}) in all calculations. 
Substituting Eq.(\ref{C}) into  Eq.(\ref{T}) one obtains
\begin{eqnarray}\label{TC}
T^2(t)=t_c^2\left(e^{-\frac{t}{t_c}}+\frac{t}{t_c}-1\right).
\end{eqnarray}
Eq.(\ref{J_kX2avr_tr}) shows that the fluctuation effect is characterized by the function 
 $\langle (\delta D)^2\rangle T^2(t)$ which takes the following form in view of Eq.(\ref{TC}): 
\begin{eqnarray}\label{T2}
\langle (\delta D)^2\rangle T^2(t) =\langle (\delta D)^2\rangle  \tau_c^2\left(e^{-\frac{t}{\tau_c}}+\frac{t}{\tau_c}-1\right).
\end{eqnarray}
If $t/\tau_c \gg 1$, then one has
\begin{eqnarray}\label{Tappr}
\langle (\delta D)^2\rangle T^2(t)\approx \langle (\delta D)^2\rangle  \tau_c t,
\end{eqnarray}
so that the fluctuations are characterized by a single dimensionless  parameter 
$\langle (\delta D)^2\rangle  \tau_c/\langle D \rangle=\frac{\langle (\delta D)^2\rangle}{\langle D \rangle^2} 
 \tau_c \langle D \rangle$ which is a product of the relative variance $\frac{\langle (\delta D)^2\rangle}{\langle D \rangle^2}$ and the dimensionless correlation time $\tau_c \langle D \rangle$. 
Another limit is $t/\tau_c\ll 1$. Then one has
\begin{eqnarray}\label{Tappr2}
\langle (\delta D)^2\rangle T^2(t)\approx \frac{1}{2} \langle (\delta D)^2\rangle   t^2,
\end{eqnarray}
so that the fluctuations are completely  characterized by the relative variance  $\langle (\delta D)^2\rangle /\langle D \rangle^2$.


\section{Numerical simulations}

\label{Section:numerics}

\begin{figure*}
   \epsfig{file=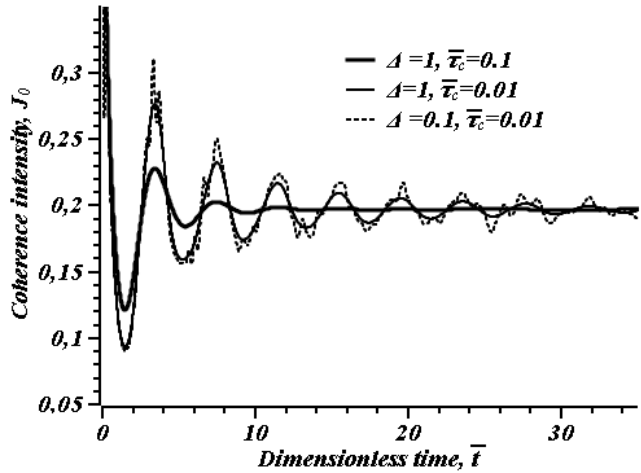
   ,scale=0.65,angle=0}
\caption{The evolution of the zero order coherence intensity $J_0(t)$ in the presence of the fluctuations with different relative variances $\Delta$ and  correlation times $\bar \tau_c$}
 \label{Fig:smooth}
\end{figure*}

 The MQ NMR dynamics of spin bearing molecules in a fluctuating nanopore can be characterized by the averaged (over the fast molecular motion in the nanopore)  dipolar coupling constant, $\langle D\rangle$, by the variance,  $\langle (\delta D)^2\rangle$ and by the correlation time, $\tau_c$, of the fluctuations. 
We use such dimensionless parameters as  the relative variance $\Delta$, the time $\bar t$ and the correlation time $\bar \tau_c$ which are introduced as follows:
\begin{eqnarray}
\Delta =  \frac{\langle (\delta D)^2\rangle}{\langle D \rangle ^2},\;\; \bar t = t \langle D \rangle,
\;\; \bar \tau_c = \tau_c \langle D \rangle.
\end{eqnarray}
First, we consider the influence of the fluctuations on the profiles of the MQ NMR coherence intensities  in  the
standard  MQ NMR experiment \cite{BMGP} with the spin-1/2 particle system of 201 equivalent spins.  
One has to emphasize that the MQ NMR coherence intensities are quickly oscillating functions in the system of equivalent spins \cite{DFFZ}. For this reason, in order to construct the profiles  of the MQ NMR coherence intensities   we consider the  intensities $J_k(t)$ (see Eq.(\ref{J1})) averaged over the sufficiently long time interval $31 < \bar t <45.5$ and find the profiles of these averaged intensities, see Ref.\cite{DFFZ}. 
Now, taking into account  the fluctuations, we observe that the fluctuations smooth out the oscillations of the coherence intensities. This is demonstrated in 
Fig.\ref{Fig:smooth}, where the evolution of zero-order coherence $J_0(\bar t)$ is represented for different relative variances $\Delta$ and  correlation times $\bar \tau_c$.  We see that, while reducing oscillations, the fluctuations do not change the mean value of the zero order coherence $J_0$. The same conclusion is valid for the higher order coherence intensities as well. 
Thus, one can expect that fluctuations do not change the profiles of the   MQ NMR coherence intensities averaged over the long  time interval. The latter conclusion is justified by the direct calculations of the MQ NMR profiles for the system of 201 spin-1/2 particles  for the  fluctuations with different relative variances and correlation times. Since profiles are not affected  by the fluctuations in the standard MQ NMR  experiment,  we do not discuss this case in more details.

\begin{figure*}
   \epsfig{file=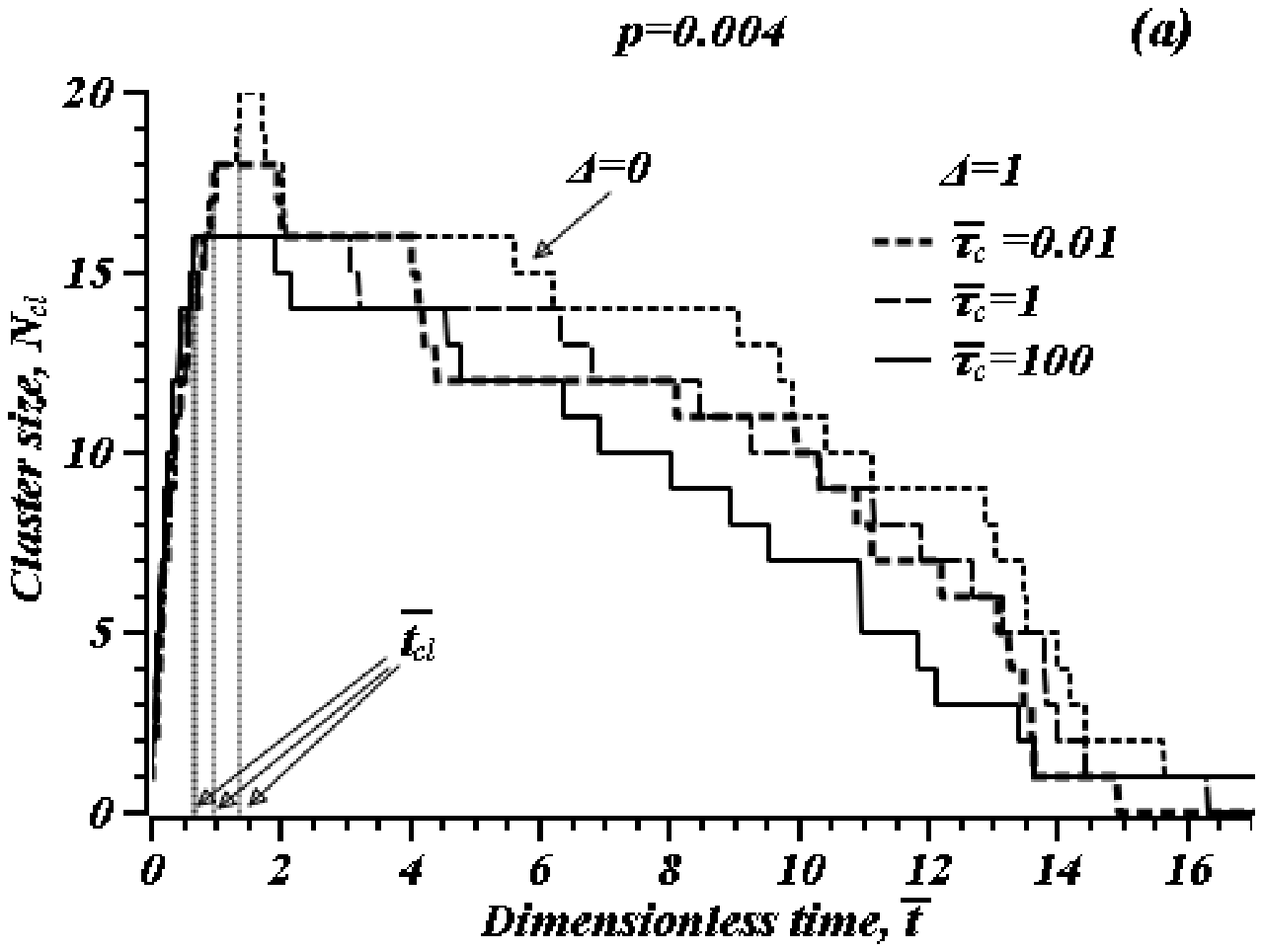
   ,scale=0.65,angle=0}
\epsfig{file=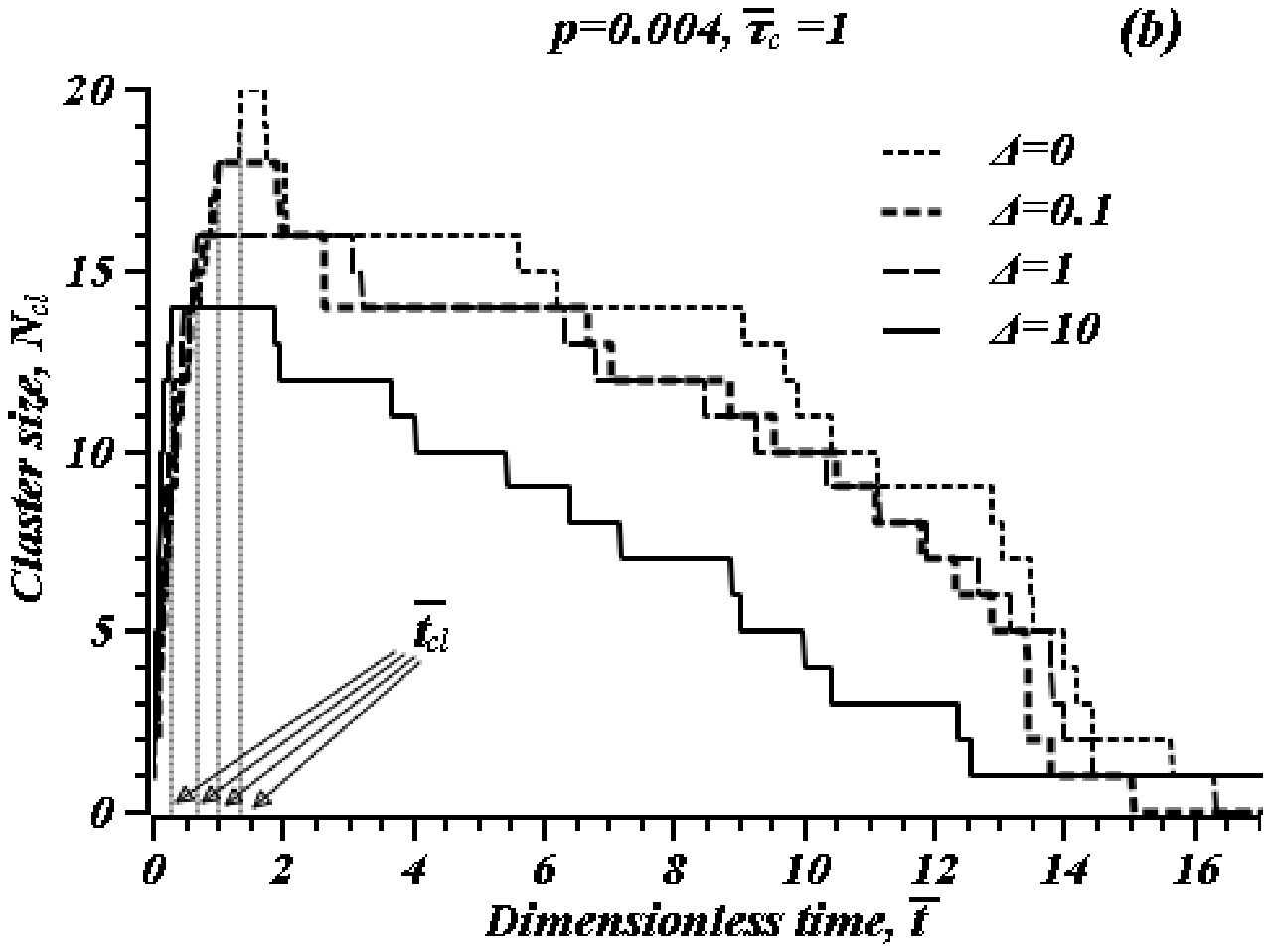
   ,scale=0.65,angle=0}
   \epsfig{file=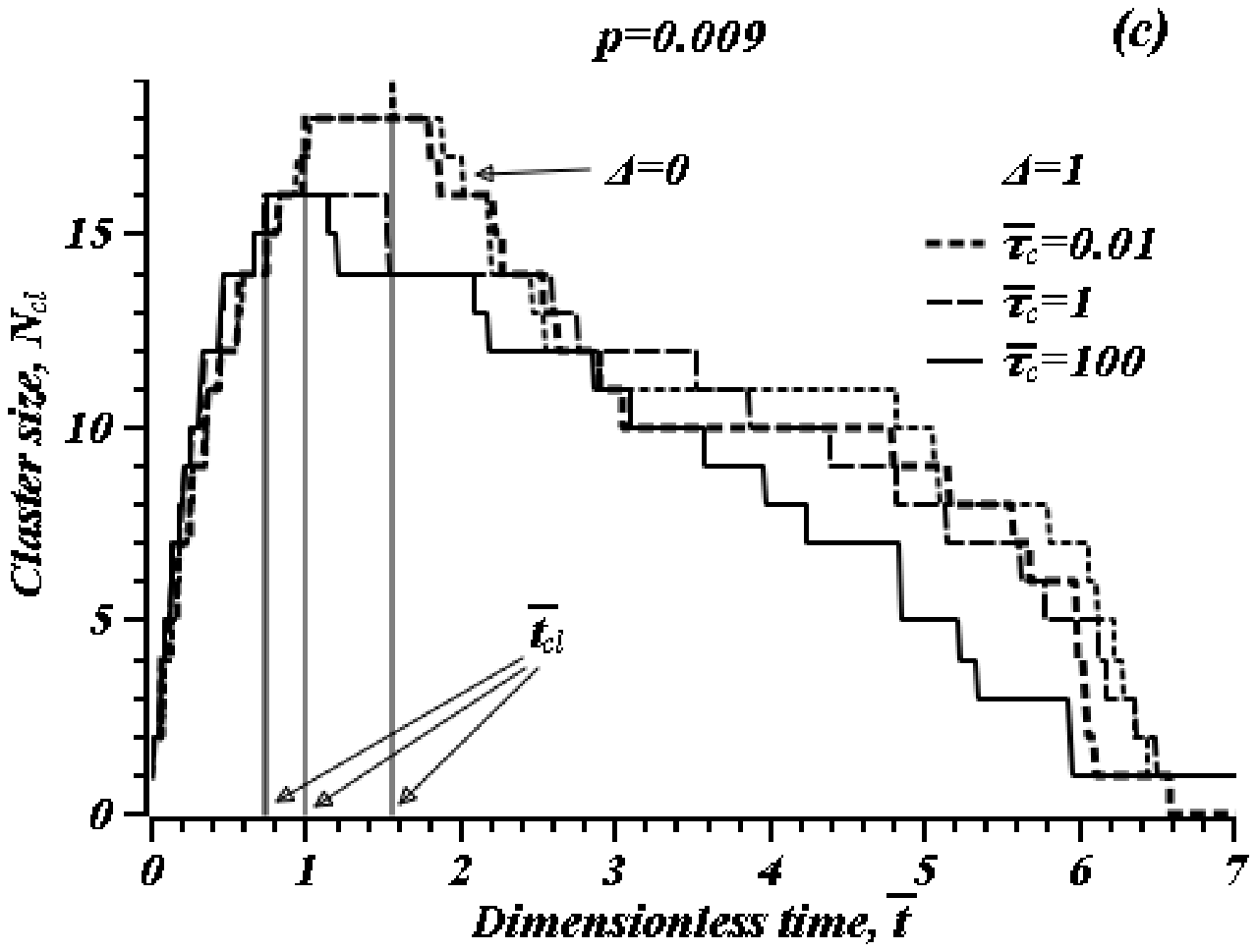
   ,scale=0.65,angle=0}
\epsfig{file=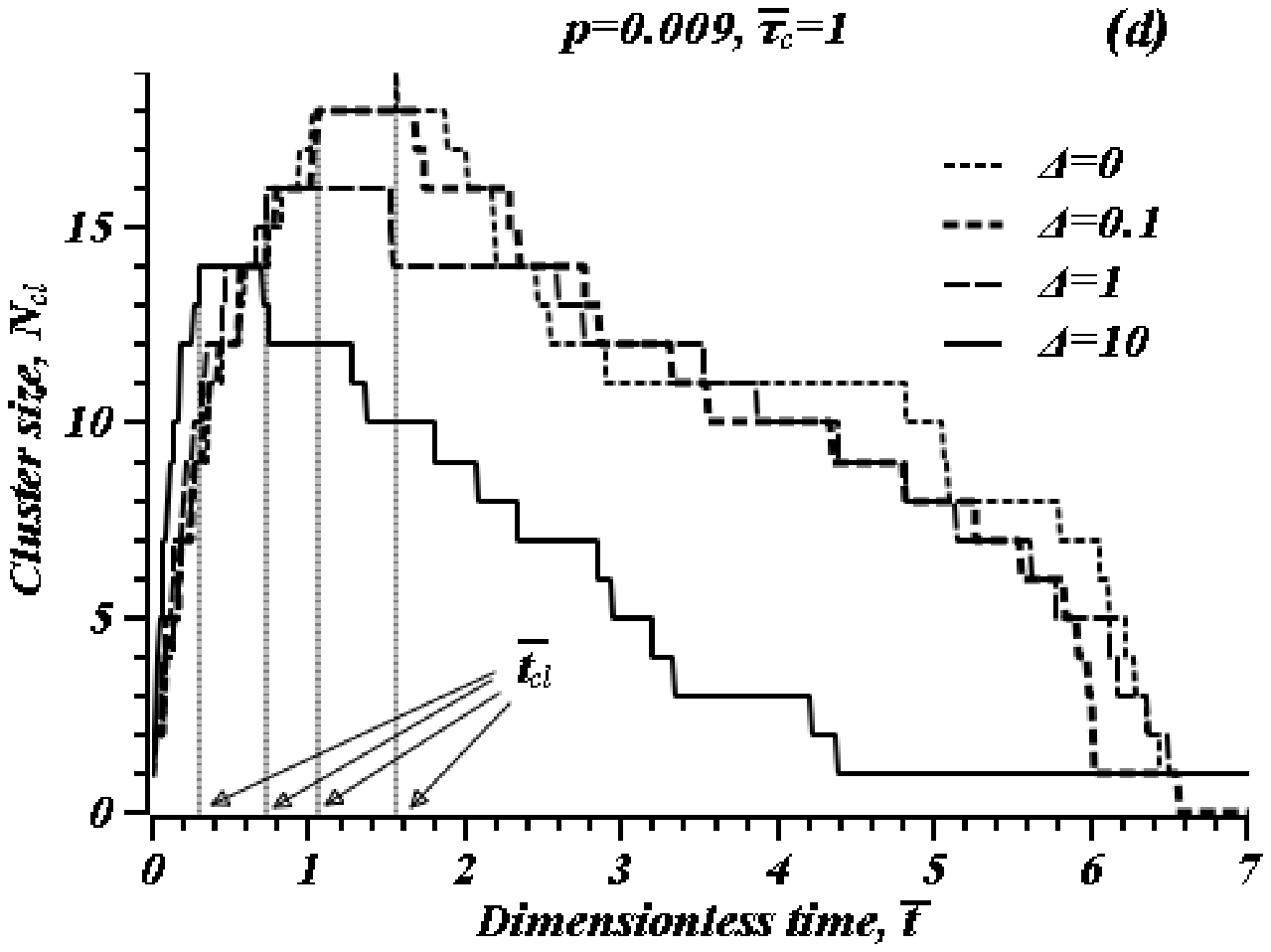
   ,scale=0.65,angle=0}
\caption{The evolution of the cluster size $N_{cl}$ under fluctuations with different relative variances
 $\Delta $  and   correlation times $\bar \tau_c$. Cluster evolution in the spin system without fluctuations ($\Delta=0$) is shown by solid lines; 
($a$) $p=0.004$, $\Delta=1$, $\bar\tau_c=0.001, \; 1,\; 100$;  
($b$) $p=0.004$, $\Delta=0.1,\; 1, \; 10$, $\bar\tau_c=1$; 
($c$) $p=0.009$, $\Delta=1$, $\bar\tau_c=0.001, \; 1,\; 100$;  
($d$) $p=0.009$, $\Delta=0.1,\; 1, \; 10$, $\bar\tau_c=1$}
\label{Fig:pp}
\end{figure*}

Now we refer to the experiments with a modified  preparation period, where the decay of MQ NMR coherences  is inherent in the preparation period \cite {AS,DFZ}. Intensities of MQ NMR coherences are given by Eq.(\ref{J2}) in  this case.  
We study the  fluctuation effect on the evolution of the coherence clusters. By the coherence cluster we mean the set of  coherences  having intensities $J_k\ge 0.005$ \cite{DFZ}.
It is obtained that the fluctuations effect both the maximal size of the coherence  cluster $N_{cl}^{max}=N_{cl}(\bar t_{cl}) $ and the  formation time of this cluster $\bar t_{cl}$. This is demonstrated  in Figs.\ref{Fig:pp},  where the evolution of the coherence  cluster size  is shown 
for two values of the parameter $p$ ($p=0.004$ in Figs.\ref{Fig:pp}($a,b$)  and $p=0.009$ in Figs.\ref{Fig:pp}($c,d$)) and different relative variances and  correlation times  of  fluctuations.
These figures demonstrate that, in general, the evolution of  the cluster size $N_{cl}$   depends on both the relative  variance $\Delta$ and the  correlation time $\bar \tau_c$. 

However, the  correlation time  effect  disappears,  if $\bar t /\bar \tau_c \ll 1$, which follows from  Eq.(\ref{Tappr2}).
This fact is suitably reflected in Table \ref{Table:D1}, where 
the maximal cluster sizes  $N_{cl}^{max}=N_{cl}(\bar t_{cl})$ and the time moments of their formation $\bar t_{cl}$ for $p=0.004$ and $p=0.009$, corresponding to the relative  variance $\Delta =1$ and different  correlation times $\bar\tau_c$ are represented. 
In fact, parameters $N_{cl}^{max}$ and $\bar t_{cl}$ corresponding to   $\bar \tau_{c}=1$ 
and  $\bar \tau_{c}=100$, are essentially the same, which is valid for both $p=0.004$ and $p=0.009$. This conclusion agrees with Eq.(\ref{Tappr2}),
 where only the relative variance is represented as a parameter responsible for the fluctuations. 

On the other hand, the relative variance  effects considerably on the cluster evolution no matter how big this variance is. In fact, 
the relative variance is present in all three formulas (\ref{T2}-\ref{Tappr2}) so that it is a significant parameter of spin dynamics. This is demonstrated in Figs.\ref{Fig:pp}($b,d$) as well as in Table \ref{Table:tc1}, where
the parameters $N_c^{max}$ and $\bar t_{cl} $ for $p=0.004$ and $p=0.009$, corresponding to the   correlation time $\bar \tau_c =1$, and different relative   variances $\Delta$  are collected. 

\begin{table}[!htb]
\begin{tabular}{|c|c|c|c||c|c|c|}
\hline
&\multicolumn{3}{|c||}{$p=0.004$}                  &\multicolumn{3}{|c|}{$p=0.009$} \\\hline
$\bar \tau_{c}$ & 0.01 & 1    & 100 &   0.01 & 1  & 100  \\\hline
$N_{cl}^{max}$                & 18   & 16   & 16  &   18   & 16 & 16   \\\hline
$\bar t_{cl} $ & 0.96 & 0.66 & 0.65&   1.00 & 0.73& 0.75  
\\\hline
\end{tabular}
\caption{  The maximal cluster  size $N_c^{max}$  and the time moment of its formation  $\bar t_{cl}$ for the relative variance $\Delta=1$
and
different   correlation times $\bar \tau_c$.  The appropriate parameters of the spin system without fluctuations are the following: $N_c^{max}=20$, 
$\bar t_{cl} =1.36$ for $p=0.04$ and  $N_c^{max}=19$, $\bar t_{cl}  =1.56$ for $p=0.09$
}
\label{Table:D1}
\end{table}

\begin{table}[!htb]
\begin{tabular}{|c|c|c|c||c|c|c|}
\hline
&\multicolumn{3}{|c||}{$p=0.004$}                  &\multicolumn{3}{|c|}{$p=0.009$} \\\hline
$\Delta$ & 0.1 & 1    & 10 &   0.1 & 1  & 10  \\\hline
$N_{cl}^{max}$                                        & 18   & 16   & 14  &   18   & 16 & 14   \\\hline
$\bar t_{cl}$                         & 0.98 & 0.66 & 0.27&   1.06 & 0.73& 0.30 
\\\hline
\end{tabular}
\caption{ The maximal cluster  size $N_c^{max}$  and the time moment of its formation 
 $\bar t_{cl}$ for the correlation time $\bar \tau_c =1$ and  different  
relative variances $\Delta$. 
}
\label{Table:tc1}
\end{table}


\section{Conclusions}

\label{Section:conclusions}

We consider fluctuations of the  dipole-dipole interaction constant $D$ in a system of equivalent spin-1/2 particles. It is found  that the fluctuations smooth out the evolution of the MQ NMR coherence intensities. This is the only fluctuation effect  unless the decay process is involved. Decay of the MQ NMR coherences has been considered in the MQ NMR experiment with the modified preparation period \cite{AS,DFZ}, where short periods of MQ coherence formation alternate with short periods of their decay.
We find that the fluctuations affect both  the maximal size of the coherence cluster and the period of its formation. Namely, 
the maximal cluster size and the period of its formation decrease  with the  increase in both  the
relative variance $\Delta$ and the  correlation time $\bar \tau_c$. However, the effect of the correlation time is reduced if $\bar t_{cl}/\bar \tau_c \ll 1$.

Although a nanopore is a rather specific 
quantum object, the numerical  results obtained for it may be useful for study of the fluctuation effects in other objects.   However, numerical simulation of the fluctuation effects in quantum systems, different from nanopores, is much more complicated because of  the exponential growth of the Hilbert space dimension   with the increase in the number of the spins, involved in the quantum process. A detailed study of different aspects of the MQ NMR dynamics \cite{DFFZ2,AS,AS2} is aimed at the increase of the power of the MQ NMR 
spectroscopy  in the investigation of the properties and structures of different physical objects. {  Using our approach, the information about volumes, shapes and orientations of  nanopores  in materials with identical nanopores can be readily extracted.   It is also possible to obtain  information about the distribution of nanopore parameters in a system with non-identical nanopores \cite{RF3}. In particular,  information about orientations of nanopores can be deduced from the free induction decay (FID) using different orientations of the sample with respect to the magnetic field. In fact, the  long time behavior of the FID is determined mainly by the nanocavities, 
where the dipolar coupling constants between spin-carring atoms (molecules) are vanishing, which corresponds to the so-called  "magic" angle orientation of the nanopore.
\cite{EBW}.} The obtained results can be also averaged over the orientations of nanopores.

All numerical simulations have been performed using the resources of the Joint Supercomputer Center (JSCC) of Russian Academy of Sciences. The authors thank Dr. S.I.Doronin for his assistance in obtaining the numerical data. The work was supported by the Program of the Presidium of Russian Academy of Sciences No.21 "Foundations of fundamental investigations of nanotechnologies and nanomaterials".


\begin{thebibliography} {150} 

\bibitem{BMGP}J.Baum, M.Munowitz, A.N.Garroway,  and A.Pines, J. Chem. Phys, {\bf  83}, 2015 (1985).

\bibitem{AS} G.A.Alvarez, and D.Suter, Phys. Rev. Lett. {\bf  104}, 230403 (2010).

\bibitem{AS2} G.A.Alvarez, and D.Suter, arXiv:1103.4546 (2011)

\bibitem{KS}
H.C.Krojanski and  D. Suter, Phys. Rev. Lett. {\bf  93}, 090501 (2004).

\bibitem{NC} M.A.Nielsen and I.L.Chuang, {\it Quantum Computation and Quantum Information}, Cambridge Univ. Press, Cambridge, 2000 



\bibitem{DFFZ} 
S.I.Doronin, A.V.Fedorova, E.B.Fel'dman and  A.I.Zenchuk,
Phys.Chem.Chem.Phys. {\bf  12}, 13273 (2010)



\bibitem{HW}
U.Haeberlen, J.S.Waugh, Phys.Rev. {\bf  175},  453 (1968)



\bibitem{FR}
E.B.Fel'dman and M.G.Rudavets, Chem.Phys.Lett. {\bf  396}, 458 (2004)





\bibitem{PWUH}
P.Poncharal, Z.L.Wang, D.Ugarte, W.A.de Heer, Science {\bf  283}, 1513 (1999)





\bibitem{SBBGH}
A.C.Sivertsen, M.J.Bajro, M.Belenky, R.G.Griffin, J.Herzfeld, Biophys. Journal {\bf  99}, 1932 (2010)



\bibitem{KNTSO}
K.Kikuchi, S.Nagata, Y.Tanoka, Y.Salhara, Z.Ogumi, J.Electroanalitical Chem. {\bf  600}, 303 (2007)

\bibitem{BKHWW} J. Baugh, A. Kleinhammes, D. Han, Q. Wang,  and Y. Wu, Science {\bf  294}, 1505 (2001).

\bibitem{FR2} E. B. Fel'dman and  M. G. Rudavets, J. Exp. Theor. Phys. {\bf  98}, 207 (2004).

\bibitem{J} J.Jeener, J.Chem.Phys. {\bf 134}, 114519 (2011)

\bibitem{J1}
J.Jeener, ''Collective effects in liquid NMR: dipolar field and radiation damping,'' in {\it Encyclopedia of Nuclear Magnetic Resonance} (John Wiley \& Sons, Ltd, Chichester 2002), Vol.9, 642



\bibitem{KN}
A.A.Khamzin, R.R.Nigmatullin, Theor.Math.Phys. {\bf 167}, 496 (2011)



\bibitem{KSN}
A.A.Khamzin, A.S.Sitdikov and A.S.Nikitin, Bull.Russ.Acad.Sci. {\bf 75}, 574 (2011)





\bibitem{DFFZ2} 
S.I.Doronin, A.V.Fedorova, E.B.Fel'dman and  A.I.Zenchuk,
J.Chem.Phys. {\bf  131} 104109 (2009)





\bibitem{DFZ} 
S.I.Doronin,  E.B.Fel'dman, and  A.I.Zenchuk,
J.Chem.Phys. {\bf  134} 034102 (2011)

\bibitem{EBW}
R.R.Ernst, G.Bodenhausen, A.Wokaun, Principles of Nuclear Magnetic Resonance in One and Two Dimensions, Clarendon, Oxford, 1987

\bibitem{A}
A.Abragam, {\it The Principles of Nuclear Magnetism}, Clarendon Press, Oxford (1961)

\bibitem{G} M. Goldman, {\it Spin Temperature and Nuclear Magnetic Resonance in Solids}, Clarendon, Oxford (1970).

\bibitem{RF3}
M.G.Rudavets, E.B.Fel'dman, Ezhegodnik Inst. Problems Chem. Phys. (in Russian), Vol.IV, 94 (2007)













\end{thebibliography}
\end{document}